# Superradiance from few emitters with non-radiative decay.

I.E.Protsenko[*,1,2], A.V.Uskov[1,3]

[1]*Lebedev Physical Institute, 119991, Leninsky prospect, 53, Moscow, Russia*

[2]*Advanced Energy Technologies ltd, 117036, Cheremuskinsky pass 5, Moscow, Russia*

[3]*ITMO University, Kronverksky pr. 49, St. Petersburg, 197101, Russia,*

[*]*Email address:protsenk@sci.lebedev.ru*

Description of superradiance of few quantum emitters with non-radiative decay in terms of quantum states is presented. Quantum efficiencies (QE) of SR of two and three emitters are calculated and compared with the case of two and three independent emitters. Maximum increase in QE is 8% for two emitters and 16% for three emitters, it is reached at certain ratios between non-radiative and radiative rates. Approach can be generalized with inclusion of the incoherent pump, dephasing and delay in emitter-emitter interaction.

*1. Introduction*

In order to consider superradiance (SR) in dissipative environment, in particular, in laser cavity with incoherent pump, one has to understand better how to describe the influence of decoherence to SR. The usual way to do it is with the density matrix formalism as, for example, in [1]. However it is much more easy to use quantum states. Description of dissipation in quantum systems in terms of quantum states is not as common as the density matrix formalism, but it has a long history. One can mention well-known Weiskopf and E. Wigner approach for spontaneous emission [2,3], its generalization to superradiance of few atoms in free space [4-6].

Here we present the way of description of superradiant quantum emitters in terms of quantum states in any dissipative environment: with non-radiative decay, incoherent pump and dephasing. One important source of decoherence is non-radiative decay of emitters, so we proceed detailed derivation of our method for particular case of superradiance with non-radiative decay. However the same approach can be applied for any other dissipation or for the incoherent pump – as we'll show. For simplicity we restrict ourselves by the case of emitters on the distance from each other much smaller than the radiation wavelength, so that we can neglect by the delay in the emitter-emitter interaction. However such delay can be easily added into consideration following [4-6]. We suppose symmetric position of emitters, when each of them equally interacts with the others, so that SR is not affected by the difference in emitter transition frequencies [7]. We'll suppose the excitation of all emitters at initial time moment and then describe their radiative and non-radiative decay.

We consider SR as cascaded or step process with radiative or non-radiative decay of only one excitation at each step. In the next two Sections 2 and 3 we describe the first and the second steps for two emitters. We find radiation power and relative quantum efficiency of SR of two emitters. Generalization of our method for many emitters became clear after description of three emitters in Sections 4 – 7. The first step for three emitters is in Section 4, the second step describing non-radiative decay – in Section 5, radiative decay – in Section 6. The rate equations for populations of states of three emitters are derived and solved in Section 7. There we calculate the radiation power and relative quantum efficiency of three superradiant emitters and compare them with the case of two emitters. Section 8 shows how to add into consideration the incoherent pump. Conclusions are presented in Section 9.

## 2. Two emitters, the first step.

We consider first two two-level emitters with the transition frequency $\omega$. They decay by emission of photons and by non-radiative decay. Each emitter has non-radiative decay to its own bath. There are various mechanisms of non-radiative decay as interactions with impurities, defects, quenching etc. Here we do not describe these mechanisms in details, but introduce some effective broadband non-radiative decay bath. The particle from this bath we call "phonon".

The state of emitter $i = 1,2$ is $|\alpha_i\rangle_i$, $\alpha_i = 1$ for exited state, $\alpha_i = 0$ for ground state. The state of two emitters is $|\alpha_1\alpha_2\rangle = |\alpha_1\rangle_1|\alpha_2\rangle_2$. There are four states of emitters: the state $|11\rangle$ with two, states $|10\rangle$, $|01\rangle$ with one excited emitter and the ground state $|00\rangle$.

States of the system are products of emitter's and bath's states. Baths states are: $|f\rangle$ of one photon in the state $f$ with certain wave vector and polarization, and $|p_i\rangle$ - the state of a phonon from non-radiative decay bath of i-th emitter. The state $|0\rangle$ is the ground bath state (no photons, no phonons). States $|f\rangle$ and $|p_i\rangle$ means that other bathes have no particles. While an emitter emits a particle to the bath, the bath particle never came back to the emitter.

The probability of decay of two emitters simultaneously is negligibly small, so we can consider the decay of emitters as a step or cascade process: only one photon or phonon is emitted in each step. We'll describe the state of emitters and baths by wave functions. At first step one of emitters comes to its ground state due to radiative or non-radiative decay, another emitter remains in its ground state. At the second step another emitter loses the excitation and come to the ground state.

The wave function of emitters and bathes for the first step is:

$$|\Psi\rangle = \left[ \begin{array}{l} C_{11}|1,1\rangle|0\rangle + \sum_f \left( C_{10,f}|1,0\rangle + C_{01,f}|0,1\rangle \right)|f\rangle + \\ \sum_{p_1} C_{01,p_1}|0,1\rangle|p_1\rangle + \sum_{p_2} C_{01,p_2}|1,0\rangle|p_2\rangle \end{array} \right] e^{-2i\omega t} \quad (1)$$

The wave function (1) shows that emitters can emit the same photon, but can't emit the same phonon. One can see in Eq.(1) entangled symmetric state of two emitters emitting photons. We come to symmetric and anti-symmetric states of emitters with one excitation and one photon:

$$|\pm\rangle|f\rangle = \frac{1}{\sqrt{2}}(|1,0\rangle \pm |0,1\rangle)|f\rangle. \quad (2)$$

Note that states (2) are orthogonal to states $|0,1\rangle|p_1\rangle$ and $|1,0\rangle|p_2\rangle$, though if we ignore bath's states and consider only emitter's states then $|\pm\rangle$ are not orthogonal to $|1,0\rangle$ and $|0,1\rangle$.

After replacement (2) the wave function (1) became

$$|\Psi\rangle = \left[ C_{11}|1,1\rangle|0\rangle + \sum_f \left( C_{+f}|+\rangle + C_{-f}|-\rangle \right)|f\rangle + \sum_{p_1} C_{01,p_1}|0,1\rangle|p_1\rangle + \sum_{p_2} C_{01,p_2}|1,0\rangle|p_2\rangle \right] e^{-2i\omega t} \quad (3)$$

The Hamiltonian describing emitter-photon and emitter-phonon interactions is:

$$H = H_0 + \sum_f \left(H_f + \hbar\hat{\Omega}_f\right) + \sum_{i=1,2}\sum_{p_i}\left(H_{p_i} + \hbar\hat{\Omega}_{p_i}\right), \tag{4}$$

where $H_0$ $H_f$ and $H_{p_i}$ describes free motion of emitters, field and phonons, $\hbar\hat{\Omega}_f$, $\hbar\hat{\Omega}_{p_i}$ describes interactions of emitters with photon and phonon baths. Equations of motion for probability amplitudes of states in the wave function (3) are:

$$\begin{aligned}
i\dot{C}_{11} &= \sum_f \frac{\Omega_f}{\sqrt{2}}\left[\left(e^{i\varphi_2} + e^{i\varphi_1}\right)C_{+,f} + \left(e^{i\varphi_2} - e^{i\varphi_1}\right)C_{-,f}\right] + \sum_{p_1}\Omega_{p_1}C_{01,p_1} + \sum_{p_2}\Omega_{p_2}C_{10,p_2} \\
i\dot{C}_{+,f} &= \delta_f C_{+,f} + \frac{\Omega_f}{\sqrt{2}}\left(e^{-i\varphi_2} + e^{-i\varphi_1}\right)C_{11} \\
i\dot{C}_{-,f} &= \delta_f C_{-,f} + \frac{\Omega_f}{\sqrt{2}}\left(e^{-i\varphi_2} - e^{-i\varphi_1}\right)C_{11} \\
i\dot{C}_{10,p_2} &= \delta_{p_2} C_{10,p_2} + \Omega_{p_2} C_{11} \\
i\dot{C}_{01,p_1} &= \delta_{p_1} C_{01,p_1} + \Omega_{p_1} C_{11}
\end{aligned} \tag{5}$$

Here $\Omega_f$ is the matrix element of transitions $|1,1\rangle|0\rangle \leftrightarrow |1,0\rangle|f\rangle$ and $|1,1\rangle|0\rangle \leftrightarrow |0,1\rangle|f\rangle$, factor $\sqrt{2}$ appears because of normalizing of states (2), $\varphi_i = \vec{k}\vec{r}_i$, $\vec{k}$ is the wave vector of emitted photon, $\vec{r}_i$ is the radius-vector of position of i-th emitter, $\delta_f = \omega_f - \omega$, $\delta_{p_i} = \omega_{p_i} - \omega$; $\omega_f$ and $\omega_{p_i}$ are photon and phonon's frequencies. Note that phase factors do not appear in terms $\sim \Omega_{p_i}$ because of each emitter has its own non-radiative "phonon" bath and phases of phonons related to different emitters are different.

We suppose that emitters are close to each other[1] so that $e^{-i\varphi_i} = 1$. It means that we neglect by the delay in the interaction between emitters through the field. Then the state $|-\rangle|f\rangle$ is not excited and

$$\begin{aligned}
i\dot{C}_{11} &= \sum_f \sqrt{2}\Omega_f C_{+,f} + \sum_{p_1}\Omega_{p_1}C_{01,p_1} + \sum_{p_2}\Omega_{p_2}C_{10,p_2} \\
i\dot{C}_{+,f} &= i\delta_f C_{+,f} + \sqrt{2}\Omega_f C_{11} \\
i\dot{C}_{10,p_2} &= i\delta_{p_2} C_{10,p_2} + \Omega_{p_2} C_{11} \\
i\dot{C}_{01,p_1} &= i\delta_{p_1} C_{01,p_1} + \Omega_{p_1} C_{11}
\end{aligned} \tag{6}$$

From Eqs.(6) we see the state $|1,1\rangle|0\rangle$ decay into three mutually orthogonal manifolds of states: $\{|+\rangle|f\rangle\}$, $\{|01\rangle|p_1\rangle\}$ and $\{|10\rangle|p_2\rangle\}$, where $f$, $p_i$ pass over many possible states of bath's particles. Supposing broad band bath spectrums and eliminating baths variables with the usual adiabatic procedure we see, that the decay rate to $\{|+\rangle|f\rangle\}$ manifold is $2\gamma_r$, where $\gamma_r$ is spontaneous radiative emission rate to free space for single emitter. There is non-radiative decay of the first and the second emitter, respectively, to $\{|01\rangle|p_1\rangle\}$ and to $\{|10\rangle|p_2\rangle\}$ manifolds, with non-radiative decay rate $\gamma$.

---

[1] One emitter is in $\vec{r}_1 = 0$

### 3. The second step

The second step does not depend on the first one, so we can suppose that initial states of the second step contains no photons and phonons. Now we have to consider the radiative and non-radiative decay of symmetric

$$|+\rangle|0\rangle = \frac{1}{\sqrt{2}}(|1,0\rangle + |0,1\rangle)|0\rangle \tag{7}$$

state to three manifolds: to the ground state of emitters with one photon $\{|00\rangle|f\rangle\}$, or to $\{|00\rangle|p_i\rangle\}$ - state with one phonon from the first or from the second emitter. State $|+\rangle|0\rangle$ decays to $\{|00\rangle|f\rangle\}$ with decay rate $2\gamma_r$, and to each of $\{|00\rangle|p_i\rangle\}$ - with the rate $\gamma/2$. Here $1/2$ appears at the procedure of adiabatic elimination of bath's variables: it is square of normalizing factor $1/\sqrt{2}$ in the wave function (7). In other words, this ½ is statistic weight of single emitter state in the entangled state (7).

Besides $|+\rangle|0\rangle$ state we have two other initial states: $|10\rangle|0\rangle$ and $|01\rangle|0\rangle$ in the second step. Single atom decays into each of these states with usual radiative and non-radiative decay rate. We can join three "ground states" $\{|00\rangle|f\rangle\}$ and $\{|00\rangle|p_i\rangle\}$ manifolds in one "ground state" manifold. Fig.1 shows all manifolds for the first and the second steps and transitions between them.

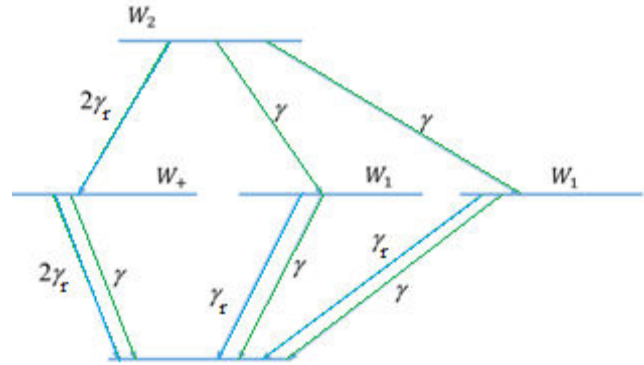

**Fig.1** Scheme of energy states of two superradiant emitters and radiative (blue arrows) and non-radiative (green arrows) transitions between them.

Dynamics of decay of emitters is described by the rate equations for populations $W_2$, $W_+$ of manifold $\{|00\rangle|f\rangle\}$ and $W_1$, which is the same for $\{|01\rangle|p_1\rangle\}$ and $\{|10\rangle|p_2\rangle\}$ manifolds:

$$\begin{aligned}\dot{W}_2 &= -2(\gamma + \gamma_r)W_2 \\ \dot{W}_+ &= 2\gamma_r W_2 - (\gamma + 2\gamma_r)W_+ \\ \dot{W}_1 &= \gamma W_2 - (\gamma + \gamma_r)W_1\end{aligned} \tag{8}$$

In order to find the radiation rate of two emitters we have to sum all rates of radiative transitions:

$$P^{(2)} = 2\gamma_r(W_2 + W_+ + W_1). \tag{9}$$

Here the upper index (2) means two emitters. At initial conditions $W_2(0)=1$, $W_+(0)=W_1(0)=0$ the solution of Eqs. (8) is:

$$W_2(t) = e^{-2(\gamma+\gamma_r)t}$$
$$W_+(t) = \frac{2\gamma_r}{\gamma}\left(1-e^{-\gamma t}\right)e^{-(2\gamma_r+\gamma)t} \qquad (10)$$
$$W_1(t) = \frac{\gamma}{\gamma+\gamma_r}\left[1-e^{-(\gamma_r+\gamma)t}\right]e^{-(\gamma_r+\gamma)t}$$

One can represent $P^{(2)}$ by noting that the radiation rate of two independent emitters is

$$P_0^{(2)}(t) = 2\gamma_r e^{-(\gamma+\gamma_r)t} = 2\gamma_r\left\{e^{-2(\gamma+\gamma_r)t}+\left[1-e^{-(\gamma_r+\gamma)t}\right]e^{-(\gamma_r+\gamma)t}\right\} \equiv 2\gamma_r\left[W_2(t)+W_1^{(0)}(t)\right], \qquad (11)$$

where $W_1^{(0)}(t)$ is the probability that one emitter is excited, while another one is not. Thus

$$P^{(2)}(t) = P_0^{(2)}(t) + 2\gamma_r\left[W_+(t) - \frac{\gamma_r}{\gamma+\gamma_r}W_1^{(0)}(t)\right], \qquad (12)$$

The second term in Eq.(12) is "addition" of radiation respectively to two independent emitters. Fig.2a shows $P^{(2)}(t)$ and $P_0^{(2)}(t)$ for $\gamma=\gamma_r$ and Fig.2b – for $\gamma=0$.

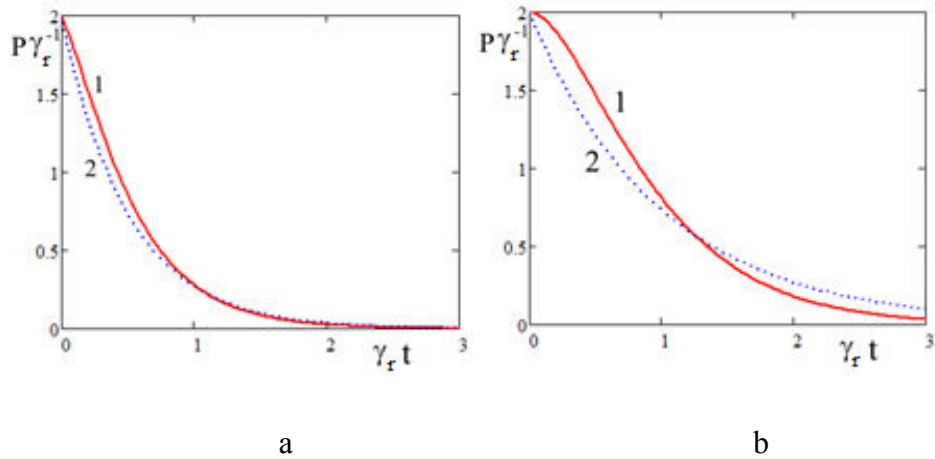

**Fig.2** Radiation power, a: for $\gamma=\gamma_r$, b: for $\gamma=0$ for two emitters with SR (curves 1) and without SR (2).

One can find total number of emitted photons (photon's yield) $Q^{(2)} = \int_0^\infty P^{(2)}(t)dt$

$$Q^{(2)} = \frac{4+2(\gamma/\gamma_r)^2+7\gamma/\gamma_r}{(\gamma/\gamma_r+1)^2(\gamma/\gamma_r+2)}, \qquad (13)$$

- with SR, and without SR:

$$Q_0^{(2)} = \frac{2}{1+\gamma/\gamma_r}. \qquad (14)$$

Obviously, that without non-radiative decay $Q^{(2)} = Q_0^{(2)} = 2$, but it is not so with non-radiative decay. Fig.3 shows the relative quantum efficiency (RQE)

$$R^{(2)} \equiv \frac{Q^{(2)}}{Q_0^{(2)}} = \frac{2+(\gamma/\gamma_r)^2 + 3.5\gamma/\gamma_r}{(\gamma/\gamma_r+1)(\gamma/\gamma_r+2)} \qquad (15)$$

of photon's yield as a function of $\gamma/\gamma_r$.

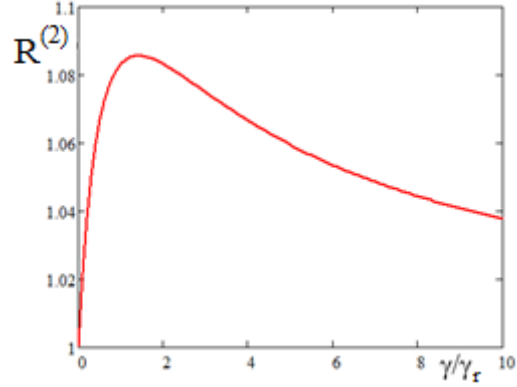

**Fig.3** Relative quantum efficiency for two superradiant and two independent emitters as function of relative non-radiative decay rate.

As one can see, $R^{(2)} > 1$: at prescience of non-radiative decay SR always increases the number of emitted photons respectively to the case of no SR. It is interesting, that at certain $\gamma/\gamma_r$ there is a maximum in the photon's yield increase. From Eq.(15) one can find $\max\left[Q^{(2)}/Q_0^{(2)}\right] = \frac{8+7\sqrt{2}}{8+6\sqrt{2}} \approx 1.086$ at $\gamma/\gamma_r = \sqrt{2} \approx 1.41$.

There is only small, about 8-9% maximum increase of efficiency for two emitters due to SR. However the acceleration of emission of two emitters by SR is also small (see Fig.2). One can expect larger acceleration of emission and increase in the number of emitted photons for more than two SR emitters, we'll see it on the example of three emitters. More than two emitters will be described the same way as two emitters: by considering decays to state manifolds including emitter's and bath's states.

4. *Three emitters The first step*

The case of three emitters is more general, than for two ones. As we'll see, it contains non-radiative relaxation transitions between symmetric Dicke states. More than three emitters can be described the same way as three emitters.

At the first step one photon or phonon is emitted from the state $|1,1,1\rangle|0\rangle$ of all three emitters excited and with no photons or phonons in baths. Energy states and transitions for the first step are shown in Fig.4

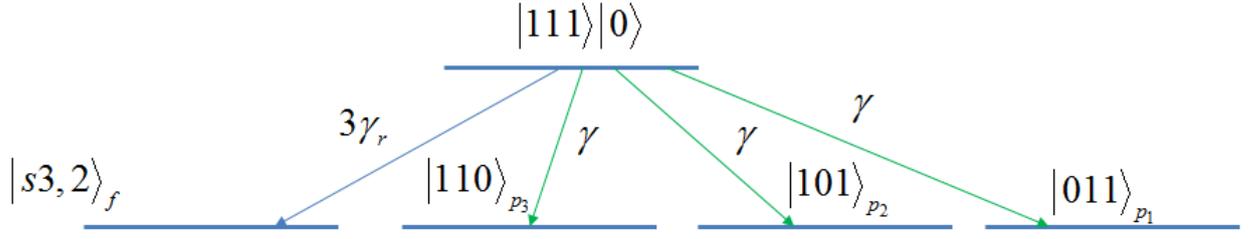

**Fig.4** Transitions from the state $|1,1,1\rangle|0\rangle$ with all three emitters excited, to states with two emitters excited and with single emitted photon or phonon.

Blue arrow in Fig.4 shows radiative transition to the manifold $|s3,2\rangle_f$, which includes entangled symmetric state $|s3,2\rangle = \left(\sqrt{3}\right)^{-1}\left(|011\rangle+|101\rangle+|011\rangle\right)$ of three emitters with two excitations and one photon $|f\rangle$; so that $|s3,2\rangle_f = |s3,2\rangle\sum_f|f\rangle$. Here $|f\rangle$ is normalized such that $\sum_f\langle f|f\rangle = 1$. The same normalizing is assumed for others baths states. Green arrows in Fig.4 indicate non-radiative relaxation of state $|1,1,1\rangle|0\rangle$ to manifolds $|011\rangle_{p1} = |011\rangle\sum_{p_1}|p_1\rangle$, $|101\rangle_{p2} = |101\rangle\sum_{p_2}|p_2\rangle$, and $|110\rangle_{p3} = |110\rangle\sum_{p_3}|p_3\rangle$. Relaxation to each manifold corresponds to transition of one of emitters to its ground state with emission of a phonon in the bath of that emitter.

5. *Next steps. Manifolds, originated from non-radiative decay.*

In the next steps one can consider manifolds, shown in Fig.4 as upper states, mutually orthogonal to each other. For simplicity in notations, we'll drop sometimes indexes, indicating presence of bath particles in upper states of the second step (for example, we'll write $|s3,2\rangle$ instead of $|s3,2\rangle_f$). We also drop a summation over states of baths appeared at the first step. But we remember that such bath particles present and make states indicated as $|s3,2\rangle$, $|011\rangle$, $|101\rangle$ and $|110\rangle$ orthogonal to each other.

Relaxation from $|011\rangle$, $|101\rangle$ and $|110\rangle$ states is going almost the same way as the relaxation of two emitters already described. We only have to remember that populations of $|011\rangle$, $|101\rangle$ and $|110\rangle$ states increases because of the relaxation from $|1,1,1\rangle|0\rangle$ state; and that the sum of populations of all states participating in the relaxation in all steps is 1.

The state $|110\rangle$, for example, decays radiatively, with the rate $2\gamma_r$, to the manifold $|s2,1\rangle_f = |s2,1\rangle_3\sum_f|f\rangle$, where $|s2,1\rangle_3 = \left(\sqrt{2}\right)^{-1}\left(|100\rangle+|010\rangle\right)$ - symmetric state of the first and the second emitter, one of them is excited, with the third emitter is in the ground state, as indicated by the index 3 in $|s2,1\rangle_3$; $|110\rangle$ decays non-radiatively to $|010\rangle_{p_1}$ and to $|100\rangle_{p_2}$ manifolds with emission of phonons to phonon's bath of the first or of the second emitter. Then, at the third step, the state $|s2,1\rangle_f$ decays radiatively, with the rate $2\gamma_r$, and non-radiatively, with the rate $\gamma$, to the ground state. States $|010\rangle_{p_1}$ and $|100\rangle_{p_2}$ decays radiatively to the ground state

with the rate $\gamma_r$ and non-radiatively with the rate $\gamma$. Transitions from $|110\rangle_{p_3}$ state to the ground state are shown in Fig.5.

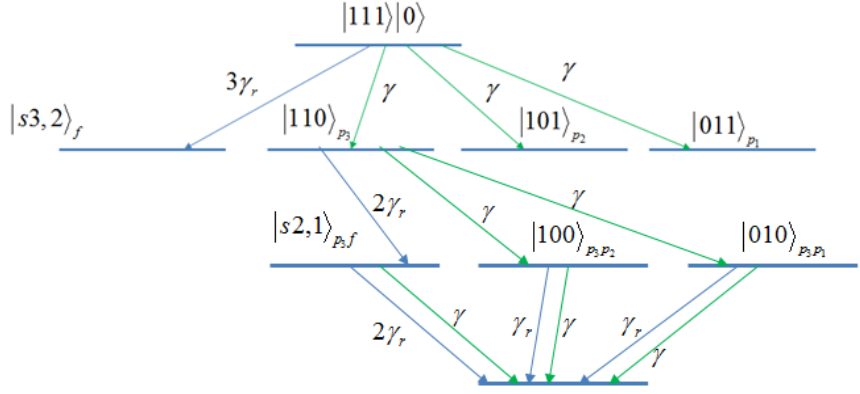

**Fig.5** Transitions from the state $|110\rangle_{p_3}$ up to the ground state shown together with transitions of the first step. Similar way transitions from $|101\rangle_{p_2}$ and from $|011\rangle_{p_1}$ states happen. Bath indexes in notations of states are preserved.

The same way radiative and non-radiative transitions happened from $|101\rangle$ and $|011\rangle$ states.

6. *The next steps. Relaxation from $|s3,2\rangle_f$ state.*

Here we drop the index $f$ in the state notation: $|s3,2\rangle_f \equiv |s3,2\rangle$ and do not show summation over states of the photon emitted at the first step. The wave function describing the relaxation from the state $|s3,2\rangle$ is:

$$C_{s3,2}|s3,2\rangle + \sum_f C_{s2,1f}|s3,1\rangle|f\rangle + \sum_{i=1}^{3}\sum_{p_i} C_{s3,1p_i}|s2,1\rangle_i|p_i\rangle \qquad (16)$$

Thus, the state $|s3,2\rangle$ radiatively decays, with the rate $4\gamma_r$, to symmetric Dicke state of three emitters (as it is in the usual Dicke model without non-radiative decay). Also $|s3,2\rangle$ decays with emission of a phonon into three symmetric Dicke states of two emitters, with the third emitter (marked by index in $|s2,1\rangle_i$ notation) in the ground state. By carrying out adiabatic elimination of phonon variable one can see that the non-radiative relaxation rate of $|s3,2\rangle$ state into any of states $|s2,1\rangle_i$ is $2\gamma/3$. The factor $2/3$ comes from formal procedure, it is statistical weight: non-radiative relaxation from two excited emitters presented in $|s3,2\rangle$ state is equally distributed between three symmetric states $|s2,1\rangle_i$ with one emitter excited. Note that each of $|s2,1\rangle_i$ state appeared at the *non-radiative* relaxation from $|s3,2\rangle$ state came with a *phonon*; while $|s2,1\rangle_i$ appeared due to *radiative* relaxation from states as $|101\rangle$, come with a *photon*, so that such symmetric states of two emitters are *orthogonal* to each other. The state $|s3,1\rangle$ decays to the ground state radiatively with the rate $3\gamma_r$ and non-radiatively with the rate $\gamma$.

The scheme of transitions for the first, second and third steps – from the state $|s3,2\rangle_f$ is shown in Fig.6.

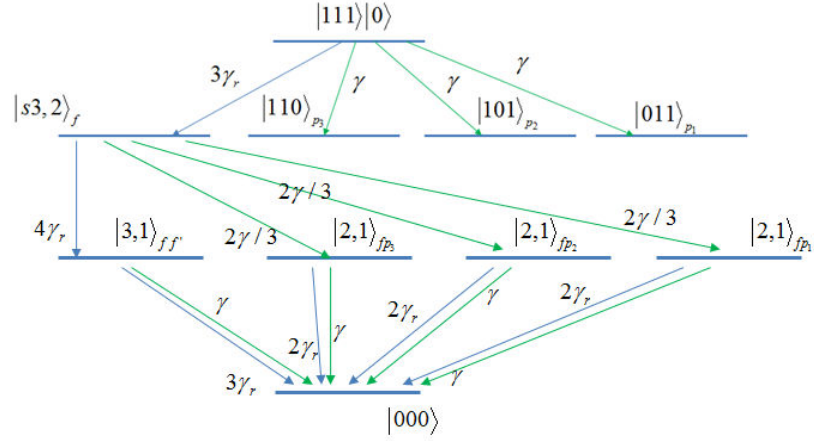

**Fig.6** Thansitions from $|s3,2\rangle_f$ state up to ground state together with transitions for the first step. Bath indexes are preserved in notations of states.

## 7. Rate equations for populations of states of three emitters and RQE

We denote $W_3$ the population of $|111\rangle$ state with all three emitters excited, $W_{3,2}^s$, $W_{3,1}^s$ - populations of symmetric Dicke states of three emitters with two ($W_{3,2}^s$) or one ($W_{3,1}^s$) of them excited; $W_{2,1}^{sr}$ $W_{2,1}^{sn}$ - population of symmetric Dicke states with two emitters, one of them is excited, for state appeared due to radiative ($W_{2,1}^{sr}$) and non-radiative ($W_{2,1}^{sn}$) transitions from an upper state. There are three states in each of these cases. Each of such states decays to the ground state through a cascade of lower states; we denote $W_2$ - populations of any of states $|011\rangle$, $|101\rangle$ and $|110\rangle$ and $W_1$ - populations of states $|001\rangle$, $|010\rangle$ and $|100\rangle$. The final set of seven equations for the state's population balance is

$$\dot{W}_3 = -3(\gamma + \gamma_r)W_3$$
$$\dot{W}_{3,2}^s = 3\gamma_r W_3 - (4\gamma_r + 2\gamma)W_{3,2}^s$$
$$\dot{W}_{3,1}^s = 4\gamma_r W_{3,2}^s - (3\gamma_r + \gamma)W_{3,1}^s$$
$$\dot{W}_{2,1}^{sn} = (2/3)\gamma W_{3,2}^s - (2\gamma_r + \gamma)W_{2,1}^{sn} \qquad (17)$$
$$\dot{W}_2 = \gamma W_3 - 2(\gamma_r + \gamma)W_2$$
$$\dot{W}_{2,1}^{sr} = 2\gamma_r W_2 - (2\gamma_r + \gamma)W_{2,1}^{sr}$$
$$\dot{W}_1 = \gamma W_2 - (\gamma_r + \gamma)W_1$$

Here first three equations are the same as in Dicke model, but with non-radiative decay. At initial condition: $W_3(0) = 1$, other populations are zero, three first ones of Eqs.(17) can be solved independently on the rest of equations. Fourth equation is for SR from symmetric state of two emitters with only one excited. The population of all stats, but $W_3$, grows because of decay from upper states. Last three equations in the set (17) describe radiative and non-radiative decays of two emitters through symmetric entangled state and through individual states of emitters. We have three states with equal populations $W_{2,1}^{sn}$ and three "cascades" of states described by the last three equations in the set (17). Each of these cascades includes two states with populations $W_1$.

Relaxation rates, include "more" $\gamma_r$-s than $\gamma$-s as, for example, $-(4\gamma_r + 2\gamma)W_{3,2}^s$ term in the second one of Eqs.(17), lead to increase of the quantum efficiency of the radiation respectively to the case of three independent emitters.

We sum all radiative relaxation (i.e. $\sim -\gamma_r$) terms in Eqs.(17) and obtain radiation power in photons per second:

$$P^{(3)}(t) = 3\gamma_r W_3 + 4\gamma_r W_{3,2}^s + 3\gamma_r W_{3,1}^s + 6\gamma_r(W_{2,1}^{sn} + W_2 + W_{2,1}^{sr} + W_1). \tag{18}$$

The first three terms in Eq.(18) are the same as in Dicke model with three emitters without non-radiative decay. The rest of Eq.(18) is the decay rate from triply degenerated states with populations $W_{2,1}^{sn}$, $W_2$ and $W_{2,1}^{sn}$, with the rate $2\gamma_r$, and 6-th degenerated state with population $W_1$ with the rate $\gamma_r$.

Solution of equations (17) can be easily found analytically starting from equation for $W_3$: $W_3(t) = e^{-3(\gamma+\gamma_r)t}$, then $W_3$ is inserted into equation for $W_{3,2}^s$ and we find $W_{3,2}^s(t) = 3\gamma_r \int_0^t W_3(t')e^{(4\gamma_r+2\gamma)(t'-t)}dt'$, etc. Fig.7a shows radiation power from three superradiant emitters calculated by formula (18) after insertion there the solution of Eqs.(17). Fig.7a shows also radiation power from three independent emitters. Fig.7b shows RQE $R^{(3)}$ for three superradiant respectively to three independent emitters: $R^{(3)} = Q^{(3)}/Q_0^{(3)}$, $Q^{(3)} = \int_0^\infty P^{(3)}(t)dt$, $Q_0^{(3)} = 3/(1+\gamma/\gamma_r)$. $R^{(3)}$ reaches the maximum max $R^{(3)} = 1.164$ at $\gamma/\gamma_r = 1.61$. Thus there is more than 16% increase of RQE, that is about two times larger than for two superradiant emitters.

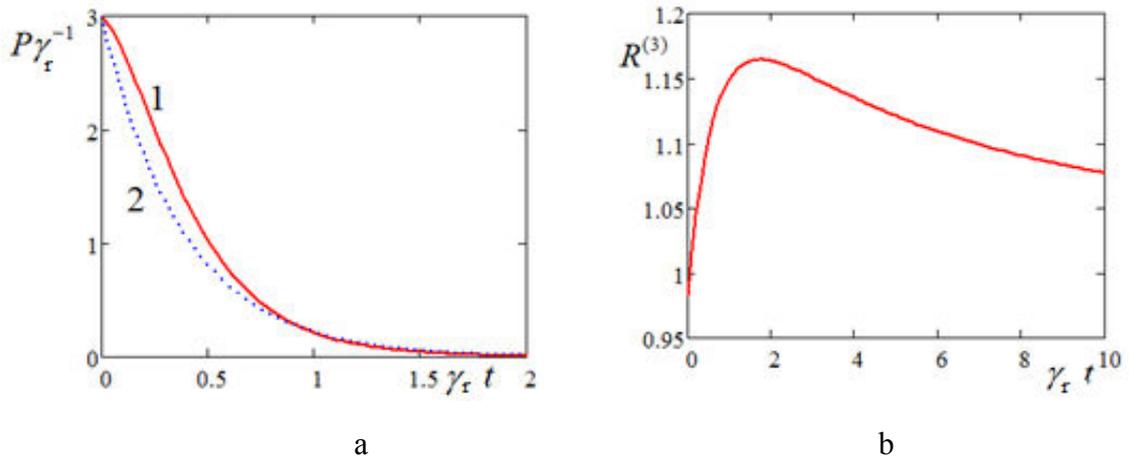

**Fig.7** (a) Radiation power in time for three superradiant emitters (curve 1) and for three independent emitters (curve 2) for $\gamma/\gamma_r = 1.61$ - when relative quantum efficiency $R^{(3)}$ has maximum. (b) $R^{(3)}$ as a function of $\gamma/\gamma_r$. $R^{(3)} = \max R^{(3)} = 1.164$ at $\gamma/\gamma_r = 1.61$.

## 8. Incoherent pump

Similar way as the non-radiative decay, one can take into consideration the incoherent pump of SR emitters - with its own pump bath for each emitter – see Fig.8. Transitions by incoherent pump for the case of two emitters are marked in Fig.8 by red arrows, each of such transition has the rate $\gamma_p$. There is no transition by incoherent pump from the ground state to the excited symmetric state with population $W_+$: the incoherent pump does not lead to appearance of symmetric state $|+\rangle$. However while the state $|+\rangle$ appears due to collective spontaneous emission from $|1,1\rangle$ state, the incoherent pump can excite $|1,1\rangle$ also from the state $|+\rangle$.

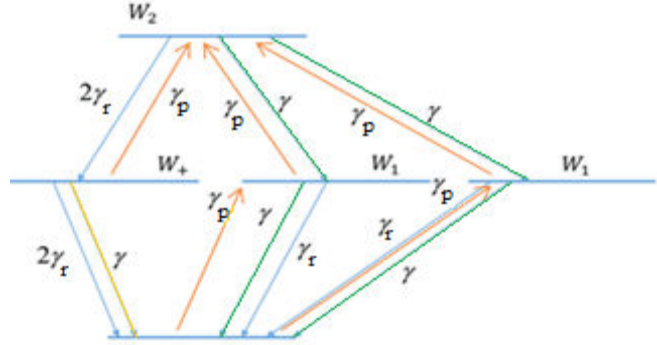

**Fig.8** Energy states of two emitters with transitions caused by incoherent pump marked by red arrows. Blue arrows mark radiative, green – non-radiative transitions.

With the incoherent pump, equations (8) of population balance became:

$$\dot{W}_2 = -2(\gamma + \gamma_r)W_2 + \gamma_p(W_+ + 2W_1)$$
$$\dot{W}_+ = 2\gamma_r W_2 - (\gamma + 2\gamma_r + \gamma_p)W_+$$
$$\dot{W}_1 = \gamma W_2 - (\gamma + \gamma_{sp} + \gamma_p)W_1 + \gamma_p W_0 \qquad (19)$$

where $W_0 = 1 - W_+ - 2W_1 - W_2$ is the ground state population. One can find the stationary solution of Eqs.(19), radiation power and RQE. Similar way one can include into consideration dephasing of emitter transitions.

## 9. Conclusion

We describe superradiance of few emitters in terms of quantum states taken into account non-radiative decay of emitters. Orthogonality between different states is provided taking into account states of the photon and phonon relaxation baths. Dynamics of emitters follows rate equations of population balance. These equation can be solved analytically or, for large number of emitters, numerically by simple iteration procedure. We considered radiation from two and three superradiant emitters and compared them with the radiation from two and three independent emitters. The case of three emitters can be obviously generalized to the case of $N > 3$ emitters. Radiation power and relative quantum efficiency (RQE) of radiation are calculated. Quantum efficiency of radiation from SR emitters is always greater than from independent emitters. Maximum RQE for two emitters is about 8%, for three emitters – about 16%: maximum RQE grows with the number of emitters. Maximum RQE is reached for certain ratio of radiative and non-radiative relaxation rates. Incoherent pump and dephasing can be taken into account the same way as the non-radiative decay. Delay in the emitter-emitter interaction can be taken into account in future studies. Results can be used for modeling SR in realistic systems with dissipation and in general, for better understanding and modeling of dynamics of quantum dissipative systems.